\documentclass[preprint,aps,amsfonts,amssymb,tightenlines,nofootinbib]{revtex4}
\usepackage{amscd}
\usepackage{amsmath}
\usepackage{eufrak}
\usepackage{eucal}
\usepackage{graphicx} 
\usepackage{bm}

\newcommand{\beq}{\begin{equation}}
\newcommand{\eeq}{\end{equation}}
\newcommand{\bea}{\begin{eqnarray}}
\newcommand{\eea}{\end{eqnarray}}

\newcommand{\nn}{\nonumber}
\newcommand{\benn}{\begin{displaymath}}
\newcommand{\eenn}{\end{displaymath}}
 
\newcommand{\tpt}{ {^3 \hskip-0.1em P_2} }

\def\pslash{p\hskip-0.45em /}

\begin{document}
\leftline{May 07 2003}
\preprint{\vbox{
\hbox{NT@UW-03-010}
\hbox{LBNL-52649}
}}

\title{
Goldstone Bosons in the $\tpt$ Superfluid Phase of Neutron Matter
and Neutrino Emission
}
\author{Paulo F. Bedaque\footnote{{\tt 
pfbedaque@lbl.gov}}, Gautam Rupak\footnote{{\tt 
grupak@lbl.gov}} }
\affiliation{Lawrence-Berkeley Laboratory, Berkeley, CA 94720}
\author{Martin J. Savage\footnote{{\tt 
savage@phys.washington.edu}}}
\affiliation{
Department of Physics, University of Washington, 
Seattle, WA 98195-1560.}

\begin{abstract}
At the high densities present in the interior of neutron stars, 
the neutrons are condensed into the $\tpt$ superfluid phase.
While this condensation has little impact on the equation of state,
it can have an important role in determining the low-temperature 
energy-momentum transport properties.
The spontaneous breaking of baryon number by the condensate
gives rise to the familiar Goldstone boson, but
in addition, the spontaneous breaking of rotational
invariance by the condensate
gives rise to three Goldstone bosons, in general, one for each
broken generator of rotations.
These Goldstone bosons, which couple to the $Z^0$, 
provide a new mechanism for neutrino emission.
Using a low-energy effective field theory to describe the dynamics of
these
Goldstone bosons we estimate the neutrino emissivity of dense neutron
matter 
and show that their annihilation
is the dominant energy-loss mechanism over a range of temperatures.
\end{abstract}
\maketitle
\bigskip

\vfill\eject

\section{Introduction}

One of the greatest challenges facing nuclear physics today 
is to understand the
behavior of nuclear matter away from nuclear matter density.
This impacts not only our understanding of the interior of neutron stars,
but
also our search for the deconfined phase of QCD in relativistic 
heavy ion collisions.
We have a good phenomenological description of the interaction between
nucleons that extends up to momenta much greater than Fermi momentum in
nuclear matter density~\cite{AV18}.  
However, when nuclear matter is significantly 
compressed this phenomenological  description becomes unreliable
as one moves toward the deconfined phase of QCD.
In fact,  a number of more ``exotic"' phases have been proposed 
as the ground state of nuclear matter at high densities,
such as the color-flavor locked (CFL) phase~\cite{CFL} 
at asymptotically high densities.

At relatively low densities we have a good
description of nuclear interactions which are dominated by the
attractive S-waves, with higher partial waves suppressed by powers
of the typical momentum.  
As the density is increased the repulsive nature of
the S-waves at higher momenta becomes important and at $\sim 1.5$ times
nuclear matter density, $\rho_{\rm nm}$,
the average $\tpt$ interactions  are the most attractive
suggesting the formation of a   $\tpt$ neutron condensate~\cite{cond}.
This has been known for sometime, and considerable work has gone into
determining the magnitude of the $\tpt$ gap as a function of density with
the most sophisticated nuclear potentials~\cite{cond,nucgap}, and also
with
effective low-energy potentials~\cite{Vlowk}.
In this work we point out that since a $\tpt$ condensate spontaneously
breaks
rotational invariance, there will be three Goldstone bosons (angulons).
In addition, baryon number is also broken, as in any superfluid, leading to the
existence of another, 
well known, Goldstone boson.
These modes will dominate the low-energy, 
low-temperature properties of the system. In particular they 
provide an important  new mechanism for neutrino emission that is not 
exponentially suppressed at temperatures below the critical temperature, 
$T_c$.

We can make a rough estimate of the size of the contribution coming from
angulon annihilation into 
a neutrino pair using dimensional analysis

\beq\label{guesstimate}
\mathcal{E}\cong G_F^2 T^9,
\eeq where $G_F$ is the Fermi constant and $T$ the temperature. This estimate assumes that powers of  dimensionful quantities like the Fermi
momentum $k_F$ or the 
value of the gap $\bar\Delta$ are not relevant. We will argue that this is indeed justified.
The temperature dependence does not have the characteristic exponential suppression
$\sim e^{-2\bar\Delta/T}$ found in processes involving gapped fermions but it
is one power of 
$T$ higher than the electron-electron scattering contribution. 
On the other hand it is not suppressed by the low electron density present in 
$\beta$-equilibrated matter. There are two caveats with the estimate in
eq.~(\ref{guesstimate}). 
First, the size of the coupling between angulons and the weak neutral gauge
boson is not at all obvious, if in fact one exists.
Second, the annihilation process is proportional to two powers of the 
angulon density $n(T)$, and 
since the Bose distribution function depends on the energy $E= v p$ of the
angulon, the density scales as  $n(T)\sim T^3/v^3$.  
If the angulon speed $v$ is small, the number of angulons in a momentum
interval is greatly enhanced, and therefore so is the emissivity. 
For this reason it is important to obtain an estimate of 
both $v$ and the dependence of the emissivity on $v$.
We will estimate these factors using an effective theory to organize our
arguments. 
A true model independent calculation is, unfortunately, not possible at the moment.

\subsection{The effective theory}

When neutrons condense in the $\tpt$ superfluid phase the
order parameter is given by
\begin{eqnarray}
i\langle  n^T  \sigma_2\sigma^j \tensor{\nabla}^k  n \rangle
& = & 
\bm{\Delta}^{jk}
\ \ ,
\label{eq:gap}
\end{eqnarray}
where $n$ are neutron field operators,
$\tensor{\nabla}^j=\overrightarrow{\nabla}^j-\overleftarrow{\nabla}^j$,
and $\sigma_2\sigma_j$ acts in spin-space.
As the neutrons are coupled together in the $\tpt$ state, the 
order parameter $\Delta^{jk}$ is a symmetric, traceless tensor.  
In general, a traceless, symmetric tensor is determined by two
orthonormal frames that diagonalize its real and imaginary parts, and two
of the eigenvalues (the third one follows from the tracelessness
condition). 
Depending on the value of the eigenvalues different phases with different unbroken symmetries
arise. It is the dynamics of the system that determines which one of those phases  is
the true ground state, symmetry arguments alone cannot determine it.
Unfortunately, not much is known about the form of the order parameter in
dense neutron matter. At temperatures close to the critical one,
arguments based on the Landau-Ginsburg energy suggest that
$\Delta_0^{ij}$ is real, but the relative sizes of the eigenvalues are
hard to predict~\cite{GL}. At zero temperature those arguments fail and much less
is known.
Due to this uncertainty we will consider here a particular choice of the
order parameter that makes some calculations feasible. 
The order of magnitude of the emissivity  likely will not depend on this choice. 

We consider the phase where the equilibrium value of the gap matrix $\Delta_0$ 
has eigenvalues equal to the cubic
roots of the unity. That is, there is
an orthonormal frame where it can be written as
\begin{eqnarray}
\Delta_0  & = & \bar\Delta
\left(
\begin{matrix}
\ 1 \ &\ 0\ &\ 0\\
 \ 0\ &\ e^{i\frac{2\pi }{3}}\ &\ 0\ \\
  \ 0\ &\ 0\ &\ e^{-i\frac{2\pi}{3}}\\ 
\end{matrix}
\right)
\ \ \ .
\label{eq:del0}
\end{eqnarray}
Rotational invariance is completely spontaneously broken down to a
discrete subgroup and as a
consequence there are three exactly massless Goldstone bosons, one for
each
rotation generator.
In addition, as $\Delta\ne 0$ spontaneously breaks baryon 
number, there is a
fourth (and well-known) Goldstone boson.
Nuclear forces do not conserve spin and orbital angular momentum
separately, due to the tensor and spin-orbit forces. The smallness
of the $\tpt$ gap, as well as its small mixing with the $^3F_2$ channel,
suggests that the effective strength of those forces at the Fermi surface
is small. For the sake of argument it will be convenient to consider the
enlarged rotation symmetry group including {\it independent} spin and
orbital rotations. This group is explicitly  broken by the tensor and spin-orbit
interactions down to the diagonal group of {\it combined} spin and orbital
rotations. The formation of the gap further  spontaneously breaks this
group down to a discrete subgroup which depends on the particular form of
the eigenvalues of $\Delta_0$.~\footnote{For special cases, as when two of
the eigenvalues are degenerate, a $O(2)$ subgroup of rotations is left
unbroken.} The analysis that follows does not depend on the particular
discrete group left unbroken and we will not discuss it further. This situation is represented by the
diagram:

\[
\begin{CD}\label{...}
{SU_S(2) \otimes SO_L(3) \otimes U(1)} @> {{\rm tensor/spin-orbit}} >>
{SO_J(3)\otimes U(1)} \\
@ VV{\langle nn\rangle\neq 0} V @ VV {\langle nn\rangle\neq 0} V \\
{\text{discrete} }@>> {{\rm tensor/spin-orbit}} > \text{discrete} \\
\end{CD}
\]
where the horizontal arrows represent explicit breaking and the vertical 
arrows denote spontaneous breaking due to the pairing.
Disregarding for the moment the  effect of the tensor and spin-orbit forces we
expect seven Goldstone bosons. 
We write the order parameter as $\bm{\Delta}= U\Delta = U\xi_S
\Delta_0 \xi_L$, where
$U=e^{i2\phi/f_0}$ is a phase and  $\xi_S$ and $\xi_L$ are orthogonal
matrices. The transformation rules under phase
($e^{i\theta}$), spin ($R_S$) or spatial ($R_L$) rotations are:
\bea
U&\rightarrow& e^{2 i\theta}\ U
\ \ \ ,\ \ \ 
\Delta \ \rightarrow \ R_S(\bm{\theta}_S )\ \Delta\ 
R_L^T(\bm{\theta}_L)
\ \ \ .
\eea
At low energies the system can be described by the most general Lagrangian 
containing the low energy degrees of freedom (the Goldstone bosons).
The Lagrangian invariant under the full $ SU_S(2) \times SO_L(3) \times
U(1)$ group is
\bea\label{eq:EFT_L}
\mathcal{L} &=& \frac{f^2}{8\bar\Delta^2}\left[ {\rm Tr} [\partial_0\Delta
\partial_0\Delta^\dagger]
 - v^2 {\rm Tr}[ \partial_i\Delta
\partial_i\Delta^\dagger]   
- w^2 \partial_i\Delta_{ik}^\dagger
\partial_j\Delta_{kj}  \right]\nn\\
&+&
\frac{f_0^2}{8} \left[ \partial_0 U  \partial_0U^\dagger -
v_0^2 \partial_iU \partial_iU^\dagger \right]  \nn\\
&+& i H_V Z_0^0 (U\partial_0 U^\dagger  -\partial_0 U U^\dagger )+
i H_A Z^0_i {\rm Tr}[ J^i(\Delta\partial_0 \Delta^\dagger  -\partial_0
\Delta \Delta^\dagger) ] \ +\  \cdots,
\eea 
where $Z_0^0$ and $Z^0_i$ are the time and spatial components of the
$Z^0$ boson. Missing from eq.~(\ref{eq:EFT_L}) are terms that break the
non-diagonal part of the rotation group, terms that do not vanish for
non-unitary $\Delta_0/\bar\Delta$, as well as terms with more derivatives whose
contribution to low energy observables are suppressed by powers of the
typical energy divided by  $\bar\Delta$.

To the terms explicitly shown in eq.~(\ref{eq:EFT_L}) we have to add
terms that break separate spin and orbital rotations.
Since the spin-orbit interaction drives the formation of the gap, its
strength is suppressed by $\sim 1/\log(\bar\Delta/\mu)$.
One of its effects is to give a mass to three out of  the seven Goldstone
bosons, specifically, to the ones corresponding to opposite spin and
orbital rotations $\xi_S = \xi_L^\dagger$. The remaining four are
strictly massless, as they correspond to the breaking of the exact
rotation and baryon number symmetry. Even though the size of this mass term is
suppressed, for small enough temperatures, the number of these pseudo-Goldstone
bosons is exponentially suppressed, 
and we will discard them.
 For the range of temperatures where it is relevant (if
any), this extra degrees of freedom could contribute to the 
$\overline{\nu}\nu$ emissivity
and our calculation should be considered as a lower bound. 
On the other
hand, the effect of the
symmetry breaking terms on the interactions should be suppressed
and we disregard them in our order of magnitude estimate.


\subsection{Determination of Low-Energy Constants}

In order for the low-energy effective field theory discussed in the
previous
section to be predictive, the {\it a priori} unknown coefficients that
enter
$v$, $w$, $f$, $H_V$ and $H_A$, must be determined from QCD.
Such a matching is not possible at this point in time 
but we are helped by the fact that all that is needed is 
information about neutron interactions close to the Fermi surface. 
Thus we can imagine integrating out modes away from the Fermi surface 
and obtaining an effective theory valid for small excitations around 
the Fermi sphere. It is known that, to leading order, 
the resulting theory is very simple and contains, besides a 
kinetic term with a modified Fermi speed, only the interactions 
leading to the formation of the gap \cite{RG_density}. 
That means that any underlying theory resulting in the same 
Fermi speed and gap will lead to the same
low energy properties of the system. We choose then a particularly simple
one
$\mathcal{L}^N = \mathcal{L}^N_S+\mathcal{L}^N_W$ with
\begin{eqnarray}
\mathcal{L}^N_S 
& = & 
n^\dagger\left( i\partial_0  +  
\frac{\nabla^2}{2 \tilde M} + \mu \right) n 
\ -\ 
g_{\tpt}  \chi_{ij}^{kl}  (n^T\sigma_2\sigma^k \tensor{\nabla}^l
n)^\dagger
 n^T\sigma_2\sigma^i \tensor{\nabla} ^j n\ ,
\nonumber\\
\mathcal{L}^N_W 
& = & 
C_V  Z_0^0\  n^\dagger n 
 +  C_A  Z_i^0\  n^\dagger\sigma^i n 
 -  g_{Z\overline{\nu}\nu}\  
Z_\mu^0 \ \overline{\nu} \gamma^\mu (1-\gamma_5)  \nu
\ \ \ ,
\label{eq:LN}
\end{eqnarray}
where the tensor
$\chi_{ij}^{kl}={1\over 2} 
(\delta_{ik}\delta_{jl}+\delta_{il}\delta_{jk}-2/3\delta_{ij}\delta_{kl})$
is a projector onto the $\tpt$ channel,
$\mu$ is the neutron chemical potential, and
$g_{\tpt}$ is an effective strong coupling constant in the $\tpt$
channel.
Not much is known about the renormalization of the weak 
interactions by the process of integrating out modes 
away from the Fermi surface. In particular we do not know whether
the form above is universal, in the same sense as the 
strong part of the Lagrangian is.
We have kept the same form of the interaction as in the vacuum and
also retained only the leading order terms in the derivative and
multi-body
expansion of the weak interactions.
The multi-body operators are
analogous to $L_{1,A}$~\cite{ButlerChen} in the pionless effective
field  theory~\cite{ChenRupakSav}.

The fact that the strong interaction in eq.~(\ref{eq:LN})
occurs in the $\tpt$ channel is suggested by the vacuum value of the
phase
shifts for nucleon-nucleon scattering, and is supported by sophisticated
nuclear models.
This is an  assumption underlying our work. 
The neutron mass is renormalized when modes far from the Fermi
surface are integrated out of the theory and thus 
$\tilde M$ is the renormalized mass or the ``in-medium'' effective mass.
An analogous renormalization occurs for the coupling to the weak 
currents and the chemical potential $\mu$.

The model in eq.~(\ref{eq:LN}) favors neutron spin-pairing and 
the formation of a gap in the $\tpt$-channel.
The strong coupling $g_{\tpt}$ in eq.~(\ref{eq:LN}) is traded for the 
neutron gap, via 
$(\tilde \Delta_0)_{ij}\ =-\ g_{\tpt}\ \langle n^T(-p) 
\ \sigma_2\sigma_i \ p_j\
n(p)\rangle$.
The tilde is to denote the model-dependence of this gap.
The propagator for neutrons in this condensed phase is
\begin{eqnarray}
\label{eq:prop}
i S(p_0, {\bf p} ) = 
\frac{i}{p_0^2-\epsilon_p^2- p^i(\tilde \Delta_0)^2_{ij}p^j}
\left(
\begin{array}{cc}
p_0+\epsilon_p  &  - i (\tilde \Delta_0)_{ij}\sigma_2\sigma_i p^j\\
   i (\tilde \Delta_0)_{ij}\sigma_i\sigma_2 p^j         & p_0-\epsilon_p 
\end{array}
\right)\ ,
\end{eqnarray}
where 
$\epsilon_p = |{\bf p}|^2/ ( 2 \tilde M) - \mu$.

The neutral current couplings in eq.~(\ref{eq:LN}) are
\begin{eqnarray}
g_{Z\overline{\nu}\nu}^2 & = & 
{G_F M_Z^2\over 2\sqrt{2} }
\ \ ,\ \ 
C_V^2\ =\ \tilde C_V^2\ {G_F M_Z^2\over 2\sqrt{2} }
\ \ ,\ \ 
C_A^2\ =\ \tilde C_A^2\ {G_F M_Z^2\over 2\sqrt{2} }
\ \ \ ,
\label{eq:weakC}
\end{eqnarray}
where $\tilde C_V=-1$, constrained by vector current conservation,
and 
$\tilde C_A=g_A+\Delta s\sim 1.1\pm 0.15$~\cite{SavWal}. 
$g_A\sim 1.26$ is the nucleon isovector axial coupling that is well
measured in nuclear $\beta$-decay, while $\Delta s$ is the 
matrix element of the strange axial-current in the proton 
that is measured in deep-inelastic scattering and
neutrino-nucleon interactions.

\subsection{Estimate of $f_0$ and $v_0$}

\begin{figure}[tb]
\begin{center}
\includegraphics[width=5in,angle=0,clip=true]{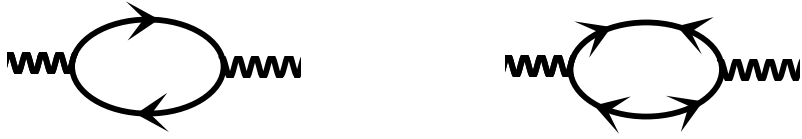}
\end{center}
\vspace*{-0pt}
\caption{
The one-loop diagrams that determine the
decay constants $f_0$ and $v_0$.
The solid line denotes a neutron, while the wiggly line denotes the
fictitious
gauge field $A_0$.
}
\label{fig:A0A0}
\end{figure}

\begin{figure}[tb]
\begin{center}
\includegraphics[width=5.5in,angle=0,clip=true]{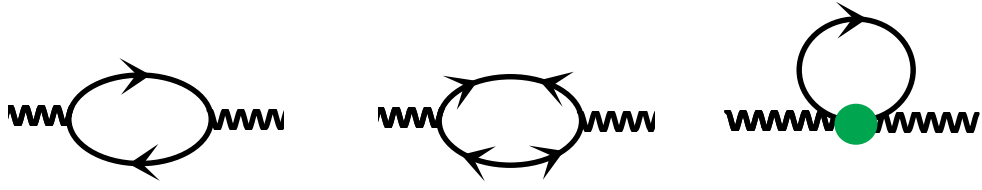}
\end{center}
\vspace*{-0pt}
\caption{
The one-loop diagrams
that determine $f$ and  the ``speeds'' $v$ and $w$ 
in eq.~(\ref{eq:EFT_L}).
The solid line denotes a neutron, while the wiggly line denotes the
fictitious
gauge field $B_i$.  The solid circle denotes an insertion of the 
tadpole vertex.
}
\label{fig:AiAi}
\end{figure}

In order to determine the parameters for the phonon $\phi$
 it is convenient to introduce a fictitious
$U(1)$ gauge-symmetry into the theory.
We then require that the low-energy effective field theory reproduce
matrix
elements of the underlying theory.
At leading order in the derivative expansion, this
amounts to replacing partial derivatives with covariant derivatives:
\begin{eqnarray}
\partial_\mu n\ \rightarrow \ D_\mu n  & = &\partial_\mu n + i A_\mu n
\ \ ,\ \ 
\partial_\mu \Delta\ \rightarrow 
\ D_\mu \Delta \ =\  \partial_\mu N +2 i A_\mu
\Delta
\ \ \ ,
\end{eqnarray}
where $A_\mu$ is the fictitious gauge field associated with the
fictitious
$U(1)$ gauge-symmetry.
In the underlying theory $A_0$ couples to the neutron 
density and acts as a chemical potential. 
The correlation function  $\langle A_0 A_0\rangle$ is the linear response
function  determining how the density changes due to a change in chemical
potential. 
In other words, it is the density of states at the Fermi surface
$dN/d\mu=Mk_F/\pi^2$ 
(up to corrections of order $\sim\Delta/\mu$), where $k_F$ is the Fermi
momentum. 
In a diagrammatic calculation, $\langle A_0 A_0\rangle$ is given 
by the two diagrams of Fig.~(\ref{fig:A0A0}). 
Using the propagator in eq.~(\ref{eq:prop}) 
we find that the two diagrams give equal contributions 
\begin{eqnarray}
 \langle A_0 A_0\rangle 
& = & 
2 \int\frac{d^4k}{(2\pi)^4} 
\frac{- (k_0+\epsilon_k)^2\ +\ 
k.
\Delta_0\Delta_0^\dagger.k}{(k_0^2-\epsilon_k^2-k.\Delta_0\Delta_0^\dagger.k)^2}
\ = \  
i \int \frac{d^3k}{(2\pi)^3} \frac{k^2 \bar\Delta^2}
 {(\epsilon_k^2+k^2 \bar\Delta^2)^\frac{3}{2}}
\nonumber\\
& \cong &  
i\ \frac{Mk_F}{\pi^2} + \mathcal{O}(\frac{{\bar \Delta^2}}{\mu^2})\ ,
\end{eqnarray}
where we used the fact that the integral is dominated by momenta
around the Fermi surface. 
Similarly, one can compute $\langle A_i A_j\rangle$.
In the underlying theory, in Coulomb gauge $\vec{\nabla}\cdot \vec{A}=0$,
it is given by the same two diagrams in
Fig.~(\ref{fig:A0A0}) plus a tadpole graph with a 
$-A_i A^i \ n^\dagger n/(2M)$ vertex, as shown in Fig.~(\ref{fig:AiAi}). 
In contrast to the calculation of
$\langle A_0 A_0\rangle$ there is  a relative minus sign for the
anomalous
graph in Fig.~(\ref{fig:AiAi}) due to the derivative coupling at the
vertices. 
Consequently, these two graphs cancel, leaving the 
tadpole contribution
\begin{eqnarray}
i\langle A_i A_j\rangle = - i \frac{N}{M}\delta_{i j}
\ \cong \ -i \frac{k_F^3}{3 \pi^2 M}\ 
\delta_{ij}
\ \ \ .
\end{eqnarray}

In the effective theory $\langle A_0 A_0\rangle$ and  $\langle A_0
A_0\rangle$ are given by the tree
level contributions and we find 
\bea
i \langle A_0 A_0\rangle &=& i \frac{M k_F}{\pi^2} = i f_0^2\ ,\nn\\
i \langle A_i A_j\rangle &=& -i \frac{k_F^3}{3\pi^2 M}\delta_{ij} = -i
f_0^2 v_0^2\delta_{ij}\ .
\eea 
We conclude that
\bea
f_0^2 &=& \frac{M k_F}{\pi^2}
\ ,\ \ \ 
v_0 \ =\  \frac{v_F^2}{3}\ .
\eea 
The breakdown of rotational
 invariance plays no role in the propagation of $\phi$
and, 
consequently, the value of $v_0$ agrees with a general analysis which assumes 
rotational invariance 
(after a suitable generalization to the non-relativistic context)~\cite{son}.


\subsection{Estimate of $f$ and $v$}

To estimate $f$ and $v$ we consider the fictitious  approximate gauge
spin rotation symmetry $n\rightarrow e^{i\frac{\bm{\sigma}}{2}.\bm{\theta}}n$.
The microscopic theory possesses this symmetry (up to corrections due to
tensor/spin-orbit forces) if the neutrons couple to the fictitious gauge
field $B_\mu$ through a covariant derivative
\beq
D_\mu n = \partial_\mu n + i B_\mu^i \frac{\sigma^i}{2}n. 
\eeq The effective theory should then be written in terms of covariant
derivatives of the $\Delta$ field
\beq
D_\mu\Delta = \partial_\mu\Delta + i B_\mu^i J^i \Delta
\ \ \ .
\eeq 
We then match the quantities $\langle B_0^i B_0^j\rangle$ and
$\langle B_k^i B_l^j\rangle$ in the microscopic and the effective theory.
The graphs contributing to them in the microscopic theory are the same
ones as in the case
of the $A$ field correlators. We find
\bea
i \langle B_0^i B_0^j\rangle &=& 
\frac{1}{4}\int \frac{d^4k}{(2\pi)^4} 
\frac{-(k_0+\epsilon_k)^2 
{\rm Tr}[ \sigma^i\sigma^j] 
- (\Delta_0.k)_a  (\Delta_0^\dagger.k)_b
{\rm Tr}[\sigma^a\sigma^i \sigma^b\sigma^j]}{(k_0^2-\epsilon_k^2)^2}\nn\\
&=& i \delta_{ij} \frac{M k_F}{6\pi^2}
\eea where we used 
${\rm Tr}[ \sigma^k\sigma^i\sigma^l\sigma^j] =
2(\delta_{ik}\delta_{jl}-\delta_{kl}\delta_{ij}+\delta_{jk}\delta_{il})$.
Notice that the denominators of the fermion propagators are spherically
symmetric, as $k.|\Delta_0|^2.k = k^2 \bar\Delta^2$. We also match the
spatial part:
\bea
i\langle B_k^i B_l^j \rangle  &=& -\frac{1}{4M^2}i^2\int
\frac{d^4k}{(2\pi)^4} 
\frac{-(k_0+\epsilon_k)^2 {\rm Tr}\sigma^i\sigma^j (-k_k k_l) + {\rm
Tr}(\Delta_0.k)_a \sigma^a\sigma^i (\Delta_0^\dagger.k)_b\sigma^b\sigma^j
k_k k_l }
{(k_0^2-E_k^2)^2} \nn\\
&-&  \frac{i}{2M} {\rm Tr}[ \sigma^i\sigma^j]  \delta_{kl}\int
\frac{d^4k}{(2\pi)^4} 
\frac{i(k_0-\epsilon_k)}
{k_0^2-\epsilon_k^2} \nn\\
&\cong& i\frac{k_F^3}{4\pi^2 M}\left[ -\frac{4}{15}\delta^{ij}\delta_{kl}
+\frac{1}{30\bar\Delta^2}
(\Delta_{0k}^i \Delta_{0l}^{\dagger j}
+\Delta_{0l}^i \Delta_{0k}^{\dagger j}
+\Delta_{0k}^j \Delta_{0l}^{\dagger i}
+\Delta_{0k}^j \Delta_{0l}^{\dagger i}) \right],
\eea 
where the integrals were computed up to corrections of orders
$\mathcal{O}(\bar\Delta/\mu)$ but the matching is valid only up to much
larger terms, of order  $g\sim 1/\log(\Delta_0/\mu)$, since the spin
rotation symmetry is only an approximate symmetry.

The same matrix elements are given in the effective theory by
\bea
i\langle B_0^i B_0^j\rangle &=& i
\frac{f^2}{2} \delta^{ij}
\ ,\ 
i\langle B_k^i B_l^j\rangle \ =\  -i \frac{
f^2 v^2}{2} \delta^{ij}\delta_{kl}-i \frac{ f^2 w^2}{8}
(\Delta_0^\dagger (J^iJ^j+J^jJ^i)\Delta_0+{\rm c.c.})_{kl}
\ ,
\eea 
and by matching to the expressions in the full theory using the relation
\beq
\left(\Delta_{0k}^i \Delta_{0l}^{\dagger j}
+\Delta_{0l}^i \Delta_{0k}^{\dagger j}
+\Delta_{0k}^j \Delta_{0l}^{\dagger i}
+\Delta_{0k}^j \Delta_{0l}^{\dagger i}  \right)=
4 {\bar\Delta^2}\delta^{ij}\delta_{kl} 
- (\Delta_0^\dagger (J^iJ^j+J^jJ^i)\Delta_0+{\rm c.c.})_{kl}
\eeq gives
\begin{eqnarray}
f^2 & = &  {Mk_F\over 3\pi^2}
\ \ ,\ \ 
v^2\ =\frac{1}{5}v_F^2
\ \ ,\ \ 
w^2\ =\frac{1}{5} v_F^2
\ \ \ .
\end{eqnarray}
We parameterize the scalars fields as $U=e^{2i\phi/f_0}$, 
$\xi=e^{i\sqrt{\frac{2}{3}}\bm{J}.\bm{\pi}/f}$ so the fields
 $\phi$ and $\pi^i$ will be canonically normalized. However, the space derivative terms
 mix the different components of $\pi^i$. The matrix for the quadratic of the 
$\pi_i$ Lagrangian reads
 \bea
 \mathcal{L}_0&=&\frac{1}{2}\!\int\!\!\frac{d^4p}{(2\pi)^4}\!\pi_i(-p)
G(p)_{i j}\pi_j(p), \\
G(p) &=&\!\!\begin{pmatrix}  p_0^2-v^2 p^2-\frac{w^2}{2}(p_y^2+p_z^2)& 
\frac{w^2}{4}p_x p_y &\frac{w^2}{4}p_x p_z\\
 \frac{w^2}{4}p_x p_y &p_0^2-v^2 p^2-\frac{w^2}{2}(p_x^2+pz^2)
 &\frac{w^2}{2}p_y p_z\\
 \frac{w^2}{4}p_x p_z & \frac{w^2}{4}p_y p_z & p_0^2-v^2
p^2-\frac{w^2}{2}(p_x^2+p_y^2)\end{pmatrix},\nonumber
\eea 
where 
$p^2=p_x^2+p_y^2+p_z^2$. 
The kinetic part can be diagonalized by using the fields $\alpha^i$ defined by
\beq
\pi_i(p)=K_{i a}(p)\alpha_a
\ \ \ .
\eeq The explicit form of $K_{i a}(p)$ can be easily found but we will omit it here since is not very enlightening.
%
\subsection{Estimate of the Weak Couplings $H_A, H_V$}
Looking at the coupling of the fictitious fields $A_\mu$ and $B_\mu$
to the neutrons we see that we can  identify
\bea
A_0 & \rightarrow & -C_V Z^0_0\ ,\nn\\
B_0^i & \rightarrow & -2 C_A (Z^0)^i
\ ,
\eea
from which we can determine $H_V$ and $H_A$ to be
\begin{eqnarray}
\label{G_match}
H_V & = & -\frac{f_0^2}{4}\ C_V
\ \ ,\ \ 
H_A \ = - \frac{f^2}{4\bar\Delta^2}\ C_A
 \ .
\end{eqnarray} 
Among other contributions, 
these weak coefficients  receive corrections of order 
$\sim 1/\log(\bar\Delta/\mu)$ coming from the strong interactions
that are {\it not} invariant under local spin rotations. 
In terms of the eigenmodes $\alpha^i$ we have the couplings
\beq
\mathcal{L}_{Z^0} = -f_0 C_V Z^0_0\partial_0\phi
-f C_A \sqrt{\frac{3}{2}} Z^0_i K_{i a}\partial_0\alpha_a
+\frac{C_A}{2}Z^0_k\epsilon^{i j k} K_{i a}(p) K_{j b}(k) 
\alpha_a(p) \partial_0\alpha_b(k)+\cdots
\eeq

\section{Neutrino emissivity}

The Goldstone bosons identified in this work will contribute to the
neutrino
emissivity of dense nuclear matter at finite temperature.
The main mechanism is the annihilation process $\alpha_i \alpha_j \rightarrow \overline{\nu}\nu$.
The amplitude for the annihilation of $\alpha_i$ and $\alpha_j$ is given
by
\begin{eqnarray}
{\cal M}_{a b} & = & 
{G_F \tilde C_A\over 4\sqrt{2}}\ 
\left[  H_{a b k} E_b(k) + H_{b a k} E_a(p)\right]\  
\overline{\nu}\gamma^k (1-\gamma_5) \nu
\ \ \ (\text{ no sum implied}),
\label{eq:abmat}
\end{eqnarray}
where  $E_i(p)$ is the energy of the angulon $\alpha_i$ with three 
momentum $p=|\bm{p}|$ and 
$H_{a b k}=\sum_{i,j=1}\epsilon^{i j k} K_{i a}(p) K_{j b}(k)$.

The emissivity due to these annihilation processes 
is defined as the energy loss due to neutrino emissions  
per unit time per unit volume, and 
for $\alpha_i \alpha_j \rightarrow \overline{\nu}\nu$ it
is given by
\begin{eqnarray}
{\cal E}_{a b} & = & 
\int 
{d^3 p_\nu\over (2\pi)^3 2 E_\nu}\ 
{d^3 p_{\overline{\nu}}\over (2\pi)^3 2 E_\nu}\ 
{d^3 p\over (2\pi)^3 2 E_a(p)}\ 
{d^3 k\over (2\pi)^3 2 E_b(k)}\ 
n(E_a(p)) \ n(E_b(k))\ 
\nonumber\\
&&
\qquad
\left( E_\nu + E_{\overline{\nu}}\right)\ 
(2\pi)^4\ \delta^{(4)} (p + k - p_\nu -p_{\overline{\nu}})
\ 
\ \sum_{s,s^\prime} |{\cal M}_{a b}|^2
\ \ \ ,
\end{eqnarray}
where the angulons $\alpha_a$, $\alpha_b$ carry incoming-momenta $p$, $k$
respectively, as indicated.  
$n(E)=[\exp(E/T)-1]^{-1}$ is the Bose distribution function and 
$\sum |{\cal M}_{a b}|^2$ is the spin summed squared matrix element
obtained from eq.~(\ref{eq:abmat}).
Using the Lorentz invariant quantity
\begin{eqnarray}
I_{\mu\nu} & = & 
\int 
{d^3 p_\nu\over (2\pi)^3 2 E_\nu}\ 
{d^3 p_{\overline{\nu}}\over (2\pi)^3 2 E_\nu}\ 
(2\pi)^4\ \delta^{(4)} (q- p_\nu -p_{\overline{\nu}})
{\rm Tr}\left[\ \pslash_\nu\gamma^\mu \pslash_{\overline{\nu}} 
\gamma_\nu (1-\gamma_5)\right]
\nonumber\\
& = & {1\over 6\pi} \left( q_\mu q_\nu - q^2 g_{\mu\nu} \right)\ 
\theta(q_0)\theta(q^2)
\ \ \ ,
\end{eqnarray}
for the neutrino phase-space integration
gives
\begin{eqnarray}
{\cal E}_{a b} & = & 
{1\over 3\pi}
\left({G_F \ \tilde C_A\over 4\sqrt{2}}\right)^2  
\int 
{d^3 p\over (2\pi)^3 2 E_a(p)}\ 
{d^3 k\over (2\pi)^3 2 E_b(k)}\ 
\left( E_a(p) + E_b(k)\right)n(E_b(p)) \ n(E_b(k))\  
\nonumber\\
&&
\qquad
\left[\ H_{a b k} E_b(k) + H_{b a k} E_a(p)\ \right]
\left[\ H_{a b l} E_b(k) + H_{b a l} E_a(p)\ \right]\nn\\
&&\quad
\left( ({\bf p}+{\bf k})_k ({\bf p}+{\bf k})_l 
+ \left(E_a(p)+E_b(k)\right)^2\delta_{kl}
- 
|{\bm p}+{\bm k}|^2\delta_{kl} \right)
\nonumber\\
&&
\quad
\ \theta\left( (E_a(p)+E_b(k))^2-|{\bm p}+{\bm k}|^2 \right)
\ \ .
\end{eqnarray} 

The phase space integral is complicated by the angular dependence of the vertex
tensor $H_{a b k}$ 
but we can 
 extract its dependence on $T$ and $v$ ($=w$). First, redefine $E_i(p)=v \tilde{E}_i(p)$
 in order to 
have the factors of $v$ explicit. Then expand the step function as
 $\theta(v^2 (\tilde{E}_i(p)+\tilde{E}_j(k))^2-|p+k|^2))\cong 
\theta(-|p+k|^2))+v^2 (\tilde{E}_i(p)+\tilde{E}_j(k))^2\delta(v^2 (\tilde{E}_i(p)+\tilde{E}_j(k))^2-|p+k|^2))+\cdots$. 
The first term does not contribute to the integral and we are left with
 \bea
 \mathcal{E} &\sim& v^5 \int 
{d^3 p\over (2\pi)^3 2 \tilde{E}_a(p)}\ 
{d^3 k\over (2\pi)^3 2 \tilde{E}_b(k)}\ 
n(v \tilde{E}_a(p)) \ n(v\tilde{E}_b(k))
  \left( \tilde{E}_a(p) + \tilde{E}_b(k)\right)^5  \nn\\
&&\quad \left[\ H_{a b k} E_b(k) + H_{b a k} E_a(p)\ \right]
\left[\ H_{a b l} E_b(k) + H_{b a l} E_a(p)\ \right]
\delta(-|p+k|^2))\nn\\
&\sim & \frac{T^9}{v^3},
 \eea where in the last step we rescale the momenta as $p\rightarrow T x/v,
 k\rightarrow T y/v$ 
and used the fact the delta function restricts the six-dimensional integral
to 
back-to-back pairs with $\vec{p}=-\vec{k}$ only. We have checked that in the 
analytically calculable case where the vertex is independent of the angle, the scaling above is obeyed.
 
Our estimate for the annihilation process is then
\bea
\mathcal{E} &=& 6\times\frac{1}{3\pi}\left( \frac{G_F \tilde C_A}{4\sqrt{2}}
\right)^2  \frac{T^9}{v^3}\nn\\
      &\cong&   10^{17}\ T_9^9 \left(\frac{0.15}{v}\right)^3\ 
\mathrm{ erg\ cm^{-3}\ s^{-s}}
\ ,
\eea where we used for the numerical estimates 
$v=0.15$ 
(corresponding to $k_F \sim 308~{\rm MeV}$, $M = 940~{\rm MeV}$, 
neglecting the
renormalization of the mass and chemical potential), $T_9=T/(10^9\ K)$ and the factor of six is 
due to the six possible combination of angulon pairs annihilated.

At typical temperatures, e.g. $T\sim 3 \times 10^8 K\sim T_c/10$, and
densities
the emissivity due to  electron bremsstrahlung is of
order $\mathcal{E}_{e}\sim 10^{10}~\mathrm{ erg\ cm^{-3}\
s^{-s}}$~\cite{Yakovlev},
which is significantly less than that due to angulon annihilation of
$\mathcal{E}_{\alpha\alpha}\sim 10^{12}~\mathrm{ erg\ cm^{-3}\
  s^{-s}}$,
where we have used $v_F\sim0.33$.
Further, comparing with processes involving the neutrons near the Fermi
surface, such as modified Urca and  neutron bremsstrahlung,
one finds that at temperatures much below the critical temperature,
$T_c$, where such processes are exponentially 
suppressed~\cite{Yakovlev,JaikumarPrakash}, the 
annihilation of angulons is likely to dominate the emissivity as it is
power-law
suppressed only.

\section{Conclusions}

We have pointed out that in the $\tpt$ neutron condensed phase that is
favored
for densities greater than $\sim 1.5 \ \rho_{\rm nm}$, 
there are Goldstone modes that contribute
to the neutrino emissivity and the energy-momentum transport properties.
Using  effective field theory arguments we estimated the
neutrino emissivity  from finite temperature, superfluid neutron matter
for a particularly simply form for the gap, one that gives a unitary order parameter.
We 
showed that the emissivity can be related to the Fermi speed and the parameters determining
the anisotropy of the gap, but is not strongly dependent on the value of the
gap itself. 
By using reasonable estimates of these parameters
we estimate an emissivity larger than other processes 
involving neutrons, which are exponentially suppressed,
and larger than that from electron bremsstrahlung, 
for the densities and temperatures relevant to neutron stars. Further,
these Goldstone bosons will likely dominate other 
low temperature observables such as neutrino opacity and  
viscosities.
Our calculation was dependent on a 
particular choice of the form 
of the gap parameter. This highlights the fact
that a determination of the actual phase of cold neutron matter would
have a larger impact on understanding the cooling processes of the $\tpt$
phase than a precise determination of the size of the gap.
A better assessment of the impact of angulon annihilation in the cooling 
of neutron stars requires the rates computed 
in this paper be inserted into a realistic cooling code.

\begin{acknowledgments}
The authors would like to acknowledge discussions with D. Kaplan and
S. Reddy, and D. Son for pointing out a flaw in a previous version of
this paper.
We also thank H. Caldas for reading the manuscript.
This work was supported by the Director, Office of Energy Research, 
Office of High Energy and Nuclear Physics, and by the Office of 
Basic Energy Sciences, Division of Nuclear Sciences, 
of the U.S. Department of Energy under Contract No.
DE-AC03-76SF00098. 
This work was also supported in
part by the U.S. Dept. of Energy under Grant No.  DE-FG03-97ER4014.
\end{acknowledgments}


\end{document}